\documentclass[12pt,onecolumn,oneside]{article}

\usepackage[cp1251]{inputenc}
\usepackage[english]{babel}
\usepackage{graphicx}
\usepackage{times}             
\usepackage{epsfig}
\usepackage[space]{cite}
\usepackage{supertabular}
\usepackage{longtable}
\usepackage{amsmath, amsfonts}
\usepackage{mathtext}

\begin{document}

\begin{center}
RADIO PULSARS WITH EXPECTED GAMMA RADIATION AND GAMMA PULSARS AS PULSATING RADIO EMITTERS
\\
 I.F. Malov and M.A. Timirkeeva
\setcounter{page}{1} 

\textit{Pushchino radioastronomical observatory, Astro space center, P.N.Lebedev physical institute, Pushchino 142290, Russia}
\\
\textit{Accepted by RAA}
 
\end{center}
\begin{abstract}
Pulsars play a crucial astrophysical role as the highly energetic compact radio, X-ray, and gamma-ray sources. Our previous works show that the radio pulsars found as the  pulsing gamma sources by the Large Area Telescope (LAT) on the board of the Fermi Gamma-Ray Space Telescope  have high values of magnetic field near the light cylinder, two-three orders of magnitude stronger comparing with the magnetic fields of radio pulsars:  $\log B_{lc}$ (G) are  3.60-3.95   and 1.75. Moreover, their losses of the rotation energy are also three orders higher than the corresponding values for the main group of radio pulsars on average:$\log \dot E$ (erg/s) = 35.37-35.53 and 32.64. The correlation between gamma-ray luminosities and radio luminosities is found. It allows us to select those objects from all set of the known radio pulsars that can be detected as gamma pulsars with the high probability. We give the list of such radio pulsars and propose to search for gamma emisson from these objects. On the other hand, the known catalog of gamma pulsars contains some sources which are not known as radio pulsars at this moment. Some of them have the large values of gamma luminosities and according to the obtained correlation, we can expect  marked radio emission from these objects.  We give the list of such pulsars and expected flux densities to search for radiation at frequencies 1400 and 111 MHz.

\end{abstract}


\section{Introduction}           
\label{sect:intro}

The space telescope Fermi with LAT gave the possibility to detect more than 150 new gamma pulsars, which are included in the Second Fermi LAT catalog (further 2FGL catalog, Abdo et al.~\cite{abdo}), and increased their previous number by a factor of several dozen. The analysis of their properties is very important for the understanding of the nature of pulsed emission. All data for gamma pulsars in our paper were taken from  this 2FGL catalog. We have used further parameters of the known radio pulsars from the ATNF catalog (Manchester et al. \cite{manchester}).

\begin{figure}
\centering
\includegraphics[width=8cm, angle=0]{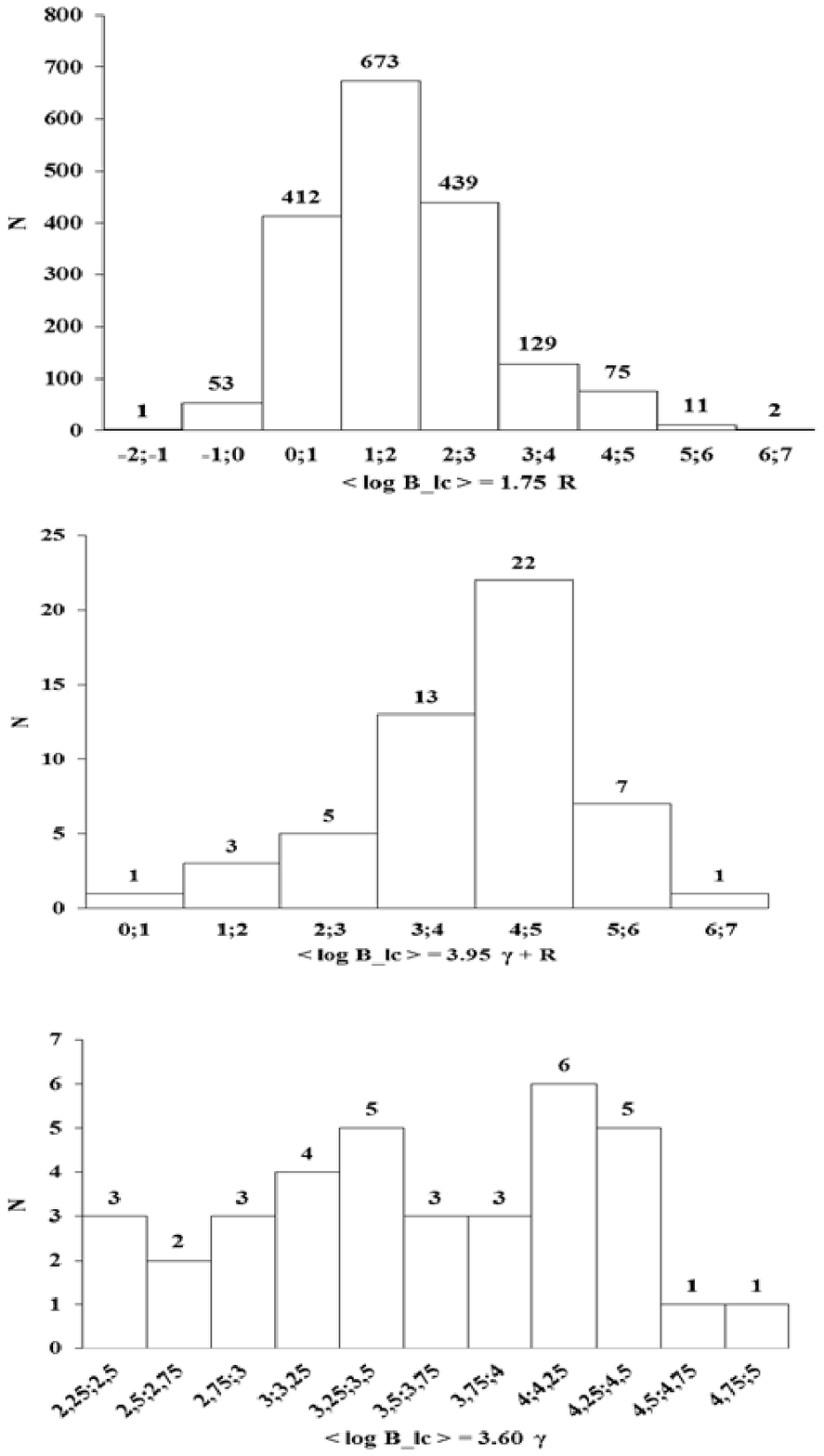}
\caption{Distributions of magnetic fields at the light cylinder for gamma quiet radio pulsars (R), gamma pulsars with radio pulsations ($\gamma$ + R) and radio quiet gamma pulsars ($\gamma$)}
\label{Fig1}
\end{figure}

\begin{figure}
\centering
\includegraphics[width=8cm, angle=0]{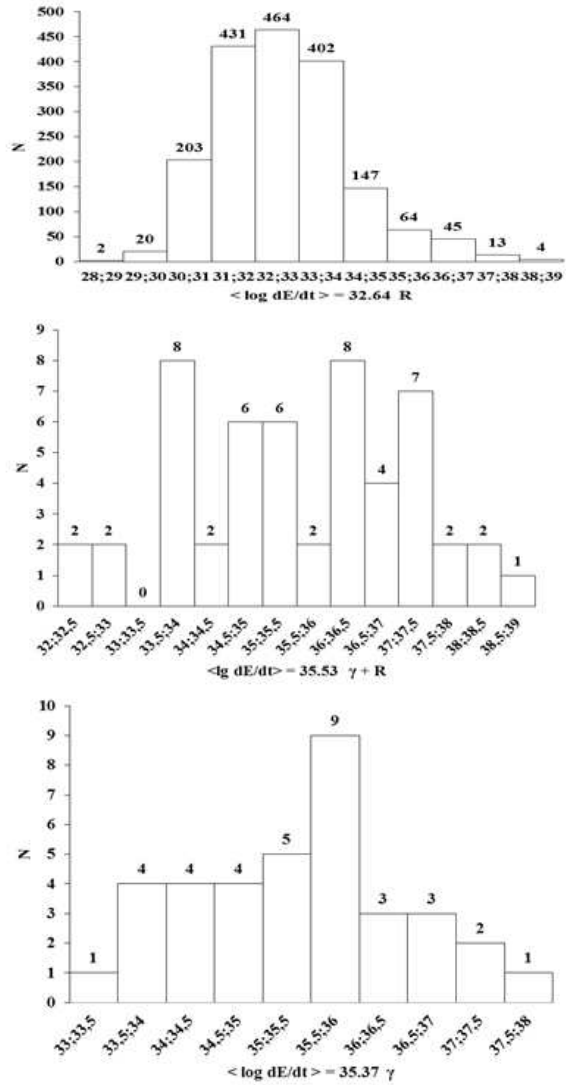}
\caption{Distributions of rotation energy losses for three samples of pulsars, as in Fig.1}
\label{Fig2}
\end{figure}

We have shown earlier (Malov \& Timirkeeva~\cite{mt1}, ~\cite{mt2}) that radio pulsars detected as pulsed  gamma sources by Fermi/LAT, are characterized by high magnetic fields at the light cylinder $B_{lc}$, two-three orders of magnitude higher than in gamma quiet radio pulsars (Fig.\ref{Fig1}). The mean values  of  $\log B_{lc}$ (G) are  3.60-3.95 and 1.75, correspondingly. Gamma pulsars  have also large values of losses of rotation energy $\dot E$, three orders of magnitude higher than the main bulk of radio pulsars (Fig.\ref{Fig2}) with the mean values of $\log \dot E$ (erg/s) = 35.37-35.53 and 32.64, correspondingly.  However, some gamma quiet radio pulsars have values of $B_{lc}$  and $\dot E$  two-three orders of magnitude higher than the means for the whole sample.  In according with our results we can expect the marked gamma emission from such  objects. On the other hand there are gamma pulsars in the Second Fermi/LAT catalog, which have not been detected as radio pulsars up to now, although they parameters imply that their radio emission must be registered. The main aim of our paper is to give the list of radio pulsars with the probable gamma radiation and show gamma pulsars with the predicted radio emission.

\section{The used sample of pulsars}
\label{sect:Sample}

We have used two criteria: magnetic fields at the light cylinder $B_{lc} > 10^3$ G and losses of rotation energy $\dot E > 3 \times 10^{34}$ erg/sec. It is worth noting that  both these parameters are determined as functions of the pulsar period and its derivative:

\begin{equation}
\dot E = \frac{4  \pi^2 I \dot P}{P^3},
\label{eq1}
\end{equation}

\begin{equation}
B_{lc} = B_s \left( \frac{R_*}{r_{lc}} \right)^3 = \frac{ 8 \pi^3  B_s R_*^3}{ c^3 P^3} = \frac{8 \pi^3 A R_*^ 3  \dot P^{1/2}}{ c^3  P^{5/2}}
\label{eq2}
\end{equation}

The last expression has been written down in the suggestion on the dipole structure of magnetic fields through the whole pulsar magnetosphere and on the pulsar braking due to the magnetic dipole radiation. Here  $I \sim 10^{45}$~ g cm$^2$ is the moment of inertia and  $R_* \sim 10^6$~  cm is the radius of the neutron star, c is the speed of light,

\begin{equation}
A = \left( \frac{3 I c^3}{ 2  \pi^2  R_*^6}
\right)^{1/2} .
\label{eq3}
\end{equation}

This means that there is the evident relation between $\dot E$ and $B_{lc}$. However, it is reasonable to take into account both these parameters, since their physical meanings differ. The losses of the rotation energy characterize the main source of pulsar energy for all processes in its magnetosphere, but the quantity  $B_{lc}$   determines the emission mechanism near the light cylinder.

\begin{figure}
\centering
\includegraphics[width=8cm, angle=0]{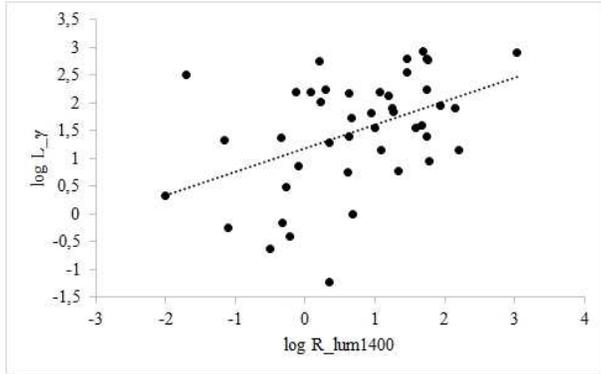}
\caption{The relationship between radio and gamma luminosities (see text)}
\label{Fig3}
\end{figure}

We exclude from the analysis pulsars in globular clusters and binary systems, since their observed characteristics can be distorted by an influence of other nearby stars, and also J1836+5925, J2021+3651, J2021+4026 and J2030+3641  with the estimated efficiency of transformation of rotation energy into gamma emission $\eta = L_{\gamma} / \dot E$ larger than $100\%$. Such high values of  $\eta$ can be caused by the suggestion on their isotropic radiation. However, the known models of gamma emission (see, for example,  Pierbattista et al. \cite{pb}) show that such emission is restricted by the rather narrow beam and the real luminosity must be lower than given by Abdo et al. \cite{abdo}. The analyzed sample is presented in Table~\ref{Tab: Tab1}. It contains values of pulsar periods $P$ (s), energy losses $\dot E$ (erg/s),  magnetic fields  $B_{lc}$ (G) at the light cylinders, radio luminosities $R_{lum1400}$ (mJy $\times$ kpc$^2$), and gamma luminosities  $L_{\gamma} (erg/s) / 10^{33}$. Comparing  radio luminosities  $R_{lum1400}$ from the catalog of ATNF Pulsar Cataloque and gamma luminosities from the 2FGL cataloq for 44 pulsars (see Table \ref{Tab: Tab1}) we obtain the following relationship (Fig.\ref{Fig3}):

\begin{equation}
\log L_{\gamma}  = (0.42 \pm 0.12) \log R_{lum1400} + 1.19  \pm 0.17,
\label{eq4}
\end{equation}

\noindent with the correlation coefficient $K = 0.45$ and the probability of the random distribution $p = 2.5 \times 10^{-3}$. Gamma luminosities in the relationship (\ref{eq4}) and in Fig.\ref{Fig3} are given as $L_{\gamma} = L (erg/s) / 10^{33}$. To obtain the eq.\ref{eq4} we have used data for all known gamma pulsars. This equation  can be used for the prediction of probable gamma emission from the known radio pulsars.


\begin{center}

\tablecaption{Considered sample of pulsars}
\tablehead{
\hline
 &    PSRJ    & P   & $R_{lum1400}$           & $\dot E$ & $B_{lc}$ & $L_{\gamma}$\\
 &            & (s) & (mJy  $\times$ kpc$^2$) &  (erg/s) & (G)      & $10^{33} (erg/s)$\\
\hline
}
\tabletail{\hline}
\begin{supertabular}{|c|c|c|c|c|c|c|}
\label{Tab: Tab1}
1	&	J0007+7303	&	0.316	&	*	&	4.5E+35	&	3.21E+03	&	94	\\
2	&	J0030+0451	&	0.005	&	0.08	&	3.5E+33	&	1.83E+04	&	0.58	\\
3	&	J0106+4855	&	0.083	&	0.07	&	2.9E+34	&	3.11E+03	&	21	\\
4	&	J0205+6449	&	0.066	&	0.46	&	2.7E+37	&	1.19E+05	&	24	\\
5	&	J0248+6021	&	0.217	&	54.8	&	2.1E+35	&	3.21E+03	&	25	\\
6	&	J0340+4130	&	0.003	&	0.79	&	7.8E+33	&	4.04E+04	&	7.3	\\
7	&	J0534+2200	&	0.033	&	56	&	4.5E+38	&	9.55E+05	&	619	\\
8	&	J0631+1036	&	0.288	&	4.15	&	1.7E+35	&	2.18E+03	&	5.6	\\
9	&	J0633+1746	&	0.237	&	*	&	3.2E+34	&	1.15E+03	&	31.7	\\
10	&	J0659+1414	&	0.385	&	0.31	&	3.8E+34	&	7.66E+02	&	0.24	\\
11	&	J0742-2822	&	0.167	&	60	&	1.4E+35	&	3.43E+03	&	9	\\
12	&	J0835-4510	&	0.089	&	86.24	&	6.9E+36	&	4.45E+04	&	89.3	\\
13	&	J0908-4913	&	0.107	&	10	&	4.9E+35	&	9.92E+03	&	35	\\
14	&	J1016-5857	&	0.107	&	4.59	&	2.6E+36	&	2.26E+04	&	55	\\
15	&	J1024-0719	&	0.005	&	2.23	&	5.3E+33	&	2.13E+04	&	0.06	\\
16	&	J1028-5819	&	0.091	&	0.73	&	8.3E+35	&	1.51E+04	&	158	\\
17	&	J1048-5832	&	0.124	&	54.66	&	2E+36	&	1.73E+04	&	176	\\
18	&	J1057-5226	&	0.197	&	*	&	3E+34	&	1.33E+03	&	4.3	\\
19	&	J1105-6107	&	0.063	&	4.18	&	2.5E+36	&	3.76E+04	&	150	\\
20	&	J1112-6103	&	0.065	&	28.35	&	4.5E+36	&	4.95E+04	&	360	\\
21	&	J1119-6127	&	0.408	&	56.45	&	2.3E+36	&	5.66E+03	&	600	\\
22	&	J1124-5916	&	0.135	&	2	&	1.2E+37	&	3.85E+04	&	170	\\
23	&	J1357-6429	&	0.166	&	4.23	&	3.1E+36	&	1.60E+04	&	25	\\
24	&	J1410-6132	&	0.050	&	1095.12	&	1E+37	&	9.58E+04	&	800	\\
25	&	J1418-6058	&	0.111	&	*	&	4.9E+36	&	3.04E+04	&	92	\\
26	&	J1420-6048	&	0.068	&	28.43	&	1E+37	&	7.13E+04	&	640	\\
27	&	J1509-5850	&	0.089	&	1.68	&	5.1E+35	&	1.22E+04	&	105	\\
28	&	J1513-5908	&	0.151	&	18.2	&	1.7E+37	&	4.17E+04	&	70	\\
29	&	J1531-5610	&	0.084	&	4.87	&	9.1E+35	&	1.71E+04	&	1	\\
30	&	J1648-4611	&	0.165	&	11.59	&	2.1E+35	&	4.18E+03	&	160	\\
31	&	J1658-5324	&	0.002	&	0.54	&	3E+34	&	1.08E+05	&	3	\\
32	&	J1702-4128	&	0.182	&	17.34	&	3.4E+35	&	4.85E+03	&	80	\\
33	&	J1709-4429	&	0.102	&	49.35	&	3.4E+36	&	2.72E+04	&	853	\\
34	&	J1718-3825	&	0.075	&	15.83	&	1.3E+36	&	2.26E+04	&	138	\\
35	&	J1730-3350	&	0.139	&	38.75	&	1.2E+36	&	1.20E+04	&	36	\\
36	&	J1732-3131	&	0.197	&	*	&	1.5E+35	&	2.93E+03	&	8.6	\\
37	&	J1741-2054	&	0.414	&	0.01	&	9.5E+33	&	3.55E+02	&	2.1	\\
38	&	J1744-1134	&	0.004	&	0.48	&	5.2E+33	&	2.68E+04	&	0.68	\\
39	&	J1747-2958	&	0.099	&	1.59	&	2.5E+36	&	2.42E+04	&	570	\\
40	&	J1747-4036	&	0.002	&	46.01	&	1.2E+35	&	3.13E+05	&	40	\\
41	&	J1801-2451	&	0.125	&	12.21	&	2.6E+36	&	1.95E+04	&	14	\\
42	&	J1809-2332	&	0.147	&	*	&	4.3E+35	&	6.74E+03	&	164	\\
43	&	J1833-1034	&	0.062	&	1.19	&	3.4E+37	&	1.42E+05	&	160	\\
44	&	J1835-1106	&	0.166	&	21.97	&	1.8E+35	&	3.84E+03	&	6	\\
45	&	J1907+0602	&	0.107	&	0.02	&	2.8E+36	&	2.38E+04	&	314	\\
46	&	J1939+2134	&	0.002	&	161.7	&	1.1E+36	&	1.02E+06	&	14	\\
47	&	J1952+3252	&	0.040	&	9	&	3.7E+36	&	7.38E+04	&	66	\\
48	&	J2043+2740	&	0.096	&	*	&	5.6E+34	&	3.73E+03	&	3.8	\\
49	&	J2124-3358	&	0.005	&	0.61	&	6.8E+33	&	2.52E+04	&	0.4	\\
50	&	J2229+6114	&	0.052	&	2.25	&	2.2E+37	&	1.39E+05	&	19.4	\\
51	&	J2240+5832	&	0.140	&	142.7	&	2.2E+35	&	5.08E+03	&	80	\\
\hline
\end{supertabular}
\end{center}

\begin{figure}
\centering
\includegraphics[width=8cm, angle=0]{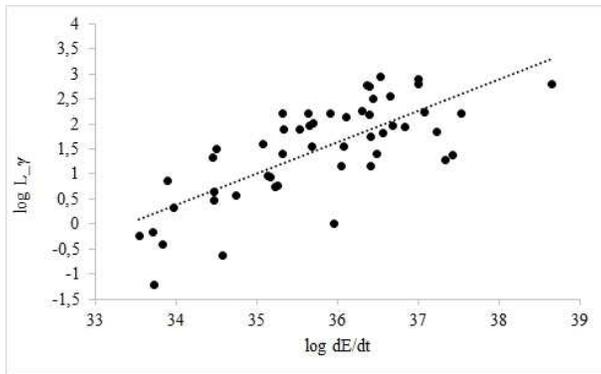}
\caption{Dependence of gamma luminosity on losses of rotation energy}
\label{Fig4}
\end{figure}

It is worth noting that the relationship (4) is the consequence of the dependence of pulsar luminosities in all ranges on the losses of their rotation energies $\dot E$. As the illustration of this conclusion Fig.\ref{Fig4} shows the dependence  $L_{\gamma} (\dot E)$ using data from Manchester et al. (\cite{manchester}) and Abdo et al. (\cite{abdo}).

The equation of the line in Fig.\ref{Fig4} has the following form:

\begin{equation}
\log L_{\gamma} (erg/s) = (0.63 \pm 0.08) \log \dot E ( erg/s) - 21.05 \pm 2.93,
\label{eq5}
\end{equation}

 \noindent the correlation coefficient $K = 0.74$ and the probability of the random distribution $p < 10^{-3}$.

  The correlation between  $\dot E$  and  $L_{\gamma}$  has been obtained for different samples of pulsars earlier by many authors (see, for example, Loginov \& Malov \cite{lm}).

\newpage    
\section{Potential gamma pulsars}
\label{sect: Gamma pulsars}
Using eq.\ref{eq4} for pulsars with $\dot E > 3 \times 10^{34}$~ erg/s and $B_{lc} > 10^3$~  G  we gave estimates of expected gamma luminosities with uncertainties in last column. These 107 pulsars (see Table \ref{Tab: Tab2}) can be detected in gamma range with the rather high probability. The certain assurance goes from the recent detections of gamma emission from the several considered objects of Table~\ref{Tab: Tab2} after the publication of the Second Fermi LAT pulsar catalog (marked as a, b and c in Table~\ref{Tab: Tab2}), the estimates made by these authors are indicated in parentheses. Their real physical estimates are numerically close to our predicted values for most objects.

\tablecaption[]{Radio pulsars with expected gamma emission}
\tablehead{
\hline
 & PSRJ    & P   & $R_{lum1400}$        & $L_{\gamma}$\\
 &         & (s) & (mJy $\times$ kpc$^2$) & $10^{33}(erg/s)$  \\	
\hline
}
\tabletail{\hline}

\begin{supertabular}{|c|c|c|c|c|}
\label{Tab: Tab2}
1	&	J0117+5914	&	0.101	&	0.94	&	15.07	$ \pm $	7.11		\\
2	&	J0358+5413	&	0.156	&	23	&	58.39	$ \pm $	45.54		\\
3	&	J0535-6935	&	0.201	&	123.5	&	119.00	$ \pm $	132.49		\\
4	&	J0538+2817	&	0.143	&	3.21	&	25.35	$ \pm $	13.22		\\
5	&	J0540-6919	&	0.051	&	59.28	&	87.20	$ \pm $	83.42		\\
6	&	J0543+2329	&	0.246	&	21.9	&	57.19	$ \pm $	44.13		\\
7	&	J0614+2229	&	0.335	&	6.66	&	34.54	$ \pm $	20.61		\\
8	&	J0729-1448	&	0.252	&	5.07	&	30.77	$ \pm $	17.39		\\
9	&	J0820-3826	&	0.125	&	20.91	&	56.08	$ \pm $	42.84		\\
10	&	J0834-4159	&	0.121	&	5.77	&	32.50	$ \pm $	18.84		\\
11	&	J0855-4644	&	0.065	&	6.52	&	34.23	$ \pm $	20.34		\\
12	&	J0940-5428	&	0.088	&	0.1	&	5.83	$ \pm $	3.80		\\
13	&	J1015-5719	&	0.140	&	6.71	&	34.65	$ \pm $	20.71		\\
14	&	J1016-5819	&	0.088	&	2.08	&	21.10	$ \pm $	10.37		\\
15	&	J1019-5749	&	0.162	&	95.05	&	106.50	$ \pm $	112.42		\\
16	&	J1020-6026	&	0.140	&	1.5	&	18.37	$ \pm $	8.77		\\
17	&	J1052-5954	&	0.181	&	1.48	&	18.27	$ \pm $	8.72		\\
18$^a$	&	J1055-6028	&	0.100	&	11.44	&	43.44	$ \pm $	29.07	(280)	\\
19	&	J1138-6207	&	0.118	&	25.33	&	60.82	$ \pm $	48.46		\\
20$^b$	&	J1151-6108	&	0.102	&	0.3	&	9.29	$ \pm $	4.88		\\
21	&	J1156-5707	&	0.288	&	1.54	&	18.58	$ \pm $	8.89		\\
22	&	J1248-6344	&	0.198	&	13.74	&	46.94	$ \pm $	32.70		\\
23	&	J1301-6305	&	0.185	&	52.86	&	83.06	$ \pm $	77.56		\\
24	&	J1327-6400	&	0.281	&	61.12	&	88.33	$ \pm $	85.06		\\
25$^b$	&	J1341-6220	&	0.193	&	301.17	&	173.59	$ \pm $	230.42		\\
26	&	J1359-6038	&	0.128	&	190	&	142.82	$ \pm $	173.31		\\
27	&	J1400-6325	&	0.031	&	12.25	&	44.71	$ \pm $	30.37		\\
28	&	J1406-6121	&	0.213	&	19.34	&	54.26	$ \pm $	40.74		\\
29	&	J1412-6145	&	0.315	&	23.83	&	59.27	$ \pm $	46.59		\\
30	&	J1413-6141	&	0.286	&	44.59	&	77.29	$ \pm $	69.60		\\
31	&	J1437-5959	&	0.062	&	5.48	&	31.80	$ \pm $	18.25		\\
32	&	J1512-5759	&	0.129	&	280.71	&	168.49	$ \pm $	220.64		\\
33	&	J1514-5925	&	0.149	&	4.15	&	28.27	$ \pm $	15.40		\\
34	&	J1524-5625	&	0.078	&	9.48	&	40.11	$ \pm $	25.78		\\
35	&	J1538-5551	&	0.105	&	8.94	&	39.13	$ \pm $	24.83		\\
36	&	J1541-5535	&	0.296	&	5.88	&	32.76	$ \pm $	19.07		\\
37	&	J1548-5607	&	0.171	&	32.6	&	67.69	$ \pm $	56.97		\\
38	&	J1601-5335	&	0.288	&	2.8	&	23.93	$ \pm $	12.22		\\
39	&	J1611-5209	&	0.182	&	10.44	&	41.79	$ \pm $	27.42		\\
40	&	J1614-5048	&	0.232	&	63.65	&	89.86	$ \pm $	87.27		\\
41	&	J1632-4757	&	0.229	&	7.06	&	35.40	$ \pm $	21.38		\\
42	&	J1636-4440	&	0.207	&	59	&	87.02	$ \pm $	83.17		\\
43	&	J1637-4553	&	0.119	&	13.02	&	45.88	$ \pm $	31.59		\\
44	&	J1637-4642	&	0.154	&	15.1	&	48.86	$ \pm $	34.74		\\
45	&	J1638-4417	&	0.118	&	30.34	&	65.66	$ \pm $	54.41		\\
46	&	J1638-4608	&	0.278	&	6.89	&	35.04	$ \pm $	21.06		\\
47	&	J1643-4505	&	0.237	&	6.34	&	33.83	$ \pm $	19.99		\\
48	&	J1646-4346	&	0.232	&	38.28	&	72.45	$ \pm $	63.14		\\
49	&	J1702-4306	&	0.216	&	6.83	&	34.91	$ \pm $	20.94		\\
50	&	J1702-4310	&	0.241	&	13.44	&	46.50	$ \pm $	32.24		\\
51	&	J1705-3950	&	0.319	&	17.65	&	52.19	$ \pm $	38.41		\\
52	&	J1715-3903	&	0.278	&	6.4	&	33.96	$ \pm $	20.10		\\
53	&	J1721-3532	&	0.280	&	232.76	&	155.64	$ \pm $	196.54		\\
54	&	J1722-3712	&	0.236	&	19.68	&	54.66	$ \pm $	41.20		\\
55	&	J1723-3659	&	0.203	&	18.38	&	53.10	$ \pm $	39.43		\\
56$^c$	&	J1739-3023	&	0.114	&	9.42	&	40.00	$ \pm $	25.67	(16.2)	\\
57	&	J1740+1000	&	0.154	&	13.92	&	47.20	$ \pm $	32.97		\\
58	&	J1743-3153	&	0.193	&	39.25	&	73.22	$ \pm $	64.16		\\
59	&	J1755-2534	&	0.234	&	3.29	&	25.62	$ \pm $	13.41		\\
60	&	J1757-2421	&	0.234	&	37.96	&	72.19	$ \pm $	62.80		\\
61	&	J1803-2137	&	0.134	&	269.1	&	165.51	$ \pm $	214.97		\\
62	&	J1809-1917	&	0.083	&	26.73	&	62.23	$ \pm $	50.16		\\
63	&	J1815-1738	&	0.198	&	5.98	&	33.00	$ \pm $	19.27		\\
64	&	J1825-1446	&	0.279	&	51.95	&	82.46	$ \pm $	76.71		\\
65	&	J1826-1334	&	0.101	&	27.37	&	62.85	$ \pm $	50.93		\\
66	&	J1828-1057	&	0.246	&	3.03	&	24.74	$ \pm $	12.79		\\
67$^c$	&	J1828-1101	&	0.072	&	65.98	&	91.24	$ \pm $	89.28	(140)	\\
68$^c$	&	J1831-0952	&	0.067	&	4.47	&	29.17	$ \pm $	16.10		\\
69	&	J1833-0827	&	0.085	&	72.9	&	95.18	$ \pm $	95.10		\\
70	&	J1835-0643	&	0.306	&	33.28	&	68.28	$ \pm $	57.73		\\
71	&	J1835-0944	&	0.145	&	7.3	&	35.91	$ \pm $	21.84		\\
72$^c$	&	J1837-0604	&	0.096	&	15.99	&	50.06	$ \pm $	36.05	(370)	\\
73	&	J1838-0453	&	0.381	&	14.73	&	48.34	$ \pm $	34.19		\\
74	&	J1838-0549	&	0.235	&	4.76	&	29.96	$ \pm $	16.73		\\
75	&	J1839-0321	&	0.239	&	16.43	&	50.63	$ \pm $	36.68		\\
76	&	J1841-0345	&	0.204	&	20	&	55.03	$ \pm $	41.63		\\
77	&	J1841-0425	&	0.186	&	50.34	&	81.36	$ \pm $	75.19		\\
78	&	J1841-0524	&	0.446	&	3.43	&	26.08	$ \pm $	13.74		\\
79$^a$	&	J1843-1113	&	0.002	&	0.16	&	7.12	$ \pm $	4.20	(5.4)	\\
80	&	J1845-0316	&	0.208	&	9	&	39.24	$ \pm $	24.94		\\
81	&	J1850-0026	&	0.167	&	79.69	&	98.84	$ \pm $	100.60		\\
82	&	J1853-0004	&	0.101	&	24.81	&	60.29	$ \pm $	47.82		\\
83	&	J1853+0056	&	0.276	&	3.1	&	24.98	$ \pm $	12.96		\\
84	&	J1856+0113	&	0.267	&	2.07	&	21.05	$ \pm $	10.35		\\
85	&	J1856+0245	&	0.081	&	23.17	&	58.57	$ \pm $	45.76		\\
86	&	J1857+0143	&	0.140	&	15.45	&	49.33	$ \pm $	35.26		\\
87	&	J1904+0800	&	0.263	&	43.16	&	76.23	$ \pm $	68.17		\\
88	&	J1907+0631	&	0.324	&	2.89	&	24.25	$ \pm $	12.44		\\
89	&	J1907+0918	&	0.226	&	19.59	&	54.55	$ \pm $	41.08		\\
90	&	J1909+0749	&	0.237	&	15.53	&	49.44	$ \pm $	35.38		\\
91	&	J1909+0912	&	0.223	&	20.27	&	55.35	$ \pm $	41.99		\\
92	&	J1913+0832	&	0.134	&	40.34	&	74.08	$ \pm $	65.29		\\
93$^a$	&	J1913+0904	&	0.163	&	2.02	&	20.84	$ \pm $	10.21	(34)	\\
94	&	J1913+1011	&	0.036	&	10.63	&	42.11	$ \pm $	27.74		\\
95	&	J1916+1225	&	0.227	&	3.96	&	27.71	$ \pm $	14.97		\\
96	&	J1917+1353	&	0.195	&	47.5	&	79.39	$ \pm $	72.46		\\
97	&	J1922+1733	&	0.236	&	33.24	&	68.25	$ \pm $	57.69		\\
98	&	J1925+1720	&	0.076	&	1.79	&	19.80	$ \pm $	9.59		\\
99	&	J1928+1746	&	0.069	&	5.26	&	31.25	$ \pm $	17.79		\\
100	&	J1930+1852	&	0.137	&	2.94	&	24.43	$ \pm $	12.57		\\
101	&	J1932+2220	&	0.144	&	142.57	&	126.46	$ \pm $	144.93		\\
102	&	J1934+2352	&	0.178	&	9.23	&	39.66	$ \pm $	25.34		\\
103	&	J1935+2025	&	0.080	&	11.15	&	42.97	$ \pm $	28.60		\\
104	&	J1938+2213	&	0.166	&	6.9	&	35.06	$ \pm $	21.08		\\
105	&	J1948+2551	&	0.197	&	47.08	&	79.09	$ \pm $	72.05		\\
106	&	J2004+3429	&	0.241	&	12.78	&	45.52	$ \pm $	31.21		\\
107	&	J2006+3102	&	0.164	&	9.82	&	40.72	$ \pm $	26.37		\\
\end{supertabular} 

Comments to Table \ref{Tab: Tab2} see References:\\
$^a$ Hou X., Smith D.A., Guillemot L. et al., \cite{hou}\\
$^b$ Smith D.A., Guillemot L., Kerr M., Ng C., Barr E., \cite{smith}\\
$^c$ Laffon H., Smith D.A., Guillemot L., \cite{laffon}

\section{Gamma pulsars with expected radio radiation}
\label{sect: Radio pulsars}

Table~\ref{Tab: Tab1} contains  some pulsars with the certain gamma emission but without detected radio radiation. Using the same data as for  eq.~\ref{eq4} we can rewrite it in the following form:

\begin{equation}
\log R_{lum 1400} = (0.59  \pm  0.17) \log L_{\gamma} + 0.13  \pm  0.34
\label{eq5a}
\end{equation}

Such potential radio emitters  are listed in Table ~\ref{Tab: Tab3}. We excluded the pulsar J1057-5226 because its value of $\dot E$ does not satisfity the criterion $\dot E > 3 \times 10^{34}$~ erg/s. Table\ref{Tab: Tab3} contains values of pulsar periods, their gamma luminosities and distances from  Abdo et al. (\cite{abdo}), expected radio luminosities calculated using eq.6 and also expected  flux densities at 1400 and 111 MHz.  $S_{1400}$  has been calculated by dividing of $R_{lum1400}$ by $d^2$. We have calculated  expected values of $S_{111}$ using the suggestion that  the  spectrum can be described by the power law:

\begin{equation}
S_{\nu} = S_0 \nu^{-\alpha}
\label{eq6}
\end{equation}

Radio spectra of the pulsars from Table ~\ref{Tab: Tab3} are not known yet and  we take the mean value of the spectral index  $\alpha = 1.5$ (Malov \& Malofeev, \cite{mm}) for all considered  objects. In the frame of these suggestions we have

\begin{equation}
S_{111} = 44.8  \times S_{1400} .
\label{eq7}
\end{equation}

\begin{table}
\caption{Gamma pulsars with expected radio emission}
\label{Tab: Tab3}
\begin{center}
\begin{tabular}{|c|c|c|c|c|c|c|c|}
\hline
		 & PSRJ &      P  & $R_{lum1400}$ & $L_{\gamma}$  & $d$  &  $S_{1400}$  & $S_{111}$ \\
 & &       (s) & mJy $\times$ kpc$^2$  &    & (kpc) &  (mJy)  & (mJy)\\						\hline
1 &	J0007+7303	&	0.316	&	8.90	&	94	&	1.4	&	4.5	&	203.4	\\
2 &	J0633+1746	&	0.237	&	5.35	&	31.7	&	0.25	&	85.6	&	3835.7	\\
3 &	J1418-6058	&	0.111	&	8.81	&	92	&	1.6	&	3.4	&	154.2	\\
4 &	J1732-3131	&	0.197	&	2.91	&	8.6	&	0.64	&	7.1	&	317.9	\\
5 &	J1809-2332	&	0.147	&	11.55	&	164	&	1.7	&	4.0	&	179.0	\\
6 &	J2043+2740	&	0.096	&	1.98	&	3.8	&	1.25	&	1.3	&	56.9	\\
\hline
\end{tabular}
\end{center}
\end{table}

These estimates show that pulsars from Table \ref{Tab: Tab3} can be registered in radio range with high probability. The most perspective gamma pulsars  in the Northern hemisphere are  J0007+7303, J0633+1746.

\section{Conclusions}
\label{sect:conclusions}

The sample of pulsars detected as radio and/or gamma emitters is presented. It restricted by objects with the losses of the rotation energy $\dot E > 3 \times 10^{34}$~ erg/s and magnetic fields at the light cylinder $B_{lc}  > 10^3$~G.

The pulsars listed in 2FGL catalog show the correlation between luminosities in radio and gamma ranges.

This correlation gives the possibility to choose radio pulsars which can be detected as gamma sources by Fermi/LAT.

Using the same correlation we propose to search for radio radiation from the several gamma pulsars believed as radio quiet  up to now. It is worth noting that  J0633+1746 has been detected already as radio pulsar (see Malofeev et al. \cite{geminga}). This detection shows that the proposed program of the search for new radio pulsars is quite reasonable.

We proposed the program of observations using LPA in Pushchino to search for radio emission from objects in Table \ref{Tab: Tab3} at 111 MHz. The results of these observations will be published separately.

\textit{This work has been carried out with the financial support of Basic Research Program of the Presidium of the Russian Academy of Sciences "Transitional and Explosive Processes in Astrophysics (P-41)" and Russian Foundation for Basic Research (grant 16-02-00954).
}

\newpage
\appendix                  

\newpage

\label{lastpage}

\end{document}